\documentclass[12pt,twoside]{article}
\usepackage{fleqn,espcrc1,epsfig}
\usepackage{graphicx}

\newcommand{\AmS}{{\protect\the\textfont2
  A\kern-.1667em\lower.5ex\hbox{M}\kern-.125emS}}
\title{
Three-quark ground-state potential in the SU(3) lattice QCD 
}

\author{H. Suganuma\address{RCNP, Osaka University, 
        Mihogaoka 10-1, Ibaraki 567-0049, Japan}
        \thanks{Present address: Tokyo Institute of Technology,
        Ohokayama 2-12-1, Meguro, Tokyo 152-8551, Japan}, 
H. Matsufuru,$^{\rm a}$ Y. Nemoto $^{\rm a}$ and T.T. Takahashi $^{\rm a}$}

\begin{document}

\maketitle

\begin{abstract}
With the smearing technique, 
the three-quark (3Q) ground-state potential $V_{\rm 3Q}$ is 
numerically extracted 
in the SU(3)$_c$ lattice QCD Monte Carlo simulation 
with $12^3 \times 24$ and $\beta=5.7$ 
at the quenched level. 
With accuracy better than a few \%, 
$V_{\rm 3Q}$ is well described by a sum of a constant $C_{\rm 3Q}$,  
the two-body Coulomb part $-A_{\rm 3Q}\sum_{i<j}
\frac1{|{\bf r}_i-{\bf r}_j|}$ and 
the three-body linear confinement part $\sigma_{\rm 3Q} L_{\rm min}$, 
where $L_{\rm min}$ denotes the minimal length of the color flux tube 
linking the three quarks. 
By comparing with the Q-$\bar {\rm Q}$ potential, 
we find a universal feature of the string tension as 
$\sigma_{\rm 3Q} \simeq \sigma_{\rm Q \bar Q}$ 
and the one-gluon-exchange result for the Coulomb coefficient as 
$A_{\rm 3Q} \simeq \frac12 A_{\rm Q \bar Q}$. 
All our results including the constant term are consistent with 
the requirement on the diquark limit in the lattice regularization.
\end{abstract}

\section{The 3Q Wilson Loop and the 3Q Ground-State Potential in QCD}

Similar to the relevant role of the Q-$\bar {\rm Q}$ potential for 
meson properties, the three-quark (3Q) 
potential \cite{TMNS99,CI86,BPV95} is directly responsible to 
the structure and properties of baryons. In spite of the importance 
of the 3Q potential in the hadron physics, there were only a few 
preliminary lattice-QCD studies for the 3Q potential done in 80's 
\cite{SW8486,TES88}. 

The 3Q ground-state potential $V_{\rm 3Q}$ 
can be measured in a gauge-invariant manner 
using the 3Q Wilson loop $W_{\rm 3Q}$ as shown in Fig.1, 
\begin{equation}
V_{\rm 3Q}=-\lim_{T \rightarrow \infty} \frac1T 
\ln \langle W_{\rm 3Q}\rangle, 
\quad 
W_{\rm 3Q} \equiv \frac1{3!}\varepsilon_{abc}\varepsilon_{a'b'c'}
U_1^{aa'} U_2^{bb'} U_3^{cc'}, 
\quad 
U_k \equiv {\rm P} e^{ig\int_{\Gamma_k}dx^\mu A_{\mu}(x)},  
\end{equation}
similar to the derivation of the Q-${\bar {\rm Q}}$ potential 
from the Wilson loop. 
In principle, $V_{\rm 3Q}$ can be obtained in the large $T$ limit, however, 
the practical lattice-QCD calculation of $\langle W_{\rm 3Q}\rangle$ 
becomes severe for large $T$, because $\langle W_{\rm 3Q}\rangle$ 
decreases exponentially with $T$. 

Physically, the true ground state of the 3Q system is expected 
to consist of three flux tubes, and 
the 3Q state expressed by the three strings generally includes 
many excited-state components such as flux-tube vibrational modes. 
Therefore, for the accurate measurement of 
the 3Q ground-state potential $V_{\rm 3Q}$, 
the ground-state enhancement or the excitation-component reduction 
by the smearing technique \cite{TMNS99,APE87,BSS95} 
is practically indispensable. 
(This smearing was not applied in Refs. \cite{SW8486,TES88}.) 

In this paper, we study the 3Q ground-state potential $V_{\rm 3Q}$ 
using the ground-state enhancement by the gauge-covariant 
smearing technique for the link-variable 
in the SU(3)$_c$ lattice QCD with the standard action with 
$\beta$=5.7 ($a \simeq $ 0.19fm) and $12^3 \times$ 24 
at the quenched level \cite{TMNS99}. 
We consider 16 patterns of the 3Q configuration where the three quarks are 
put on $(i,0,0)$, $(0,j,0)$ and $(0,0,k)$ in ${\bf R}^3$ with 
$0 \le i,j,k \le 3$ in the lattice unit. 
The junction point $O$, which does not affect $V_{\rm 3Q}$, 
is set at the origin $(0,0,0)$ in ${\bf R}^3$. 

\section{The Smearing Technique and the Ground-State Enhancement}

The smearing technique is actually successful for accurate measurements 
of the Q-$\bar {\rm Q}$ potential in the lattice QCD \cite{BSS95}.  
The standard smearing for link-variables is performed by the 
iterative replacement of the spatial link-variable $U_i (s)$ 
($i=1,2,3$) by the obscured link-variable 
$\bar U_i (s) \in {\rm SU(3)}_c$ \cite{APE87,BSS95} which maximizes 
\vspace{-0.2cm}
\begin{equation}
{\rm Re} \,\, {\rm tr} \left\{ 
\bar U_i^{\dagger}(s) [
\alpha U_i(s)+\sum_{j \ne i} \sum_{\pm} 
U_{\pm j}(s)U_i(s \pm \hat j)U_{\pm j}^\dagger (s + \hat i) 
] \right\}, \quad U_{-\mu}(s) \equiv U_{\mu}^{\dagger}(s-\hat \mu), 
\end{equation}
\vspace{-0.3cm}
with a real smearing parameter $\alpha$. 
The $n$-th smeared link-variable $U_\mu^{(n)}(s)$ 
is defined as 
\begin{equation}
U_i^{(n)}(s) \equiv \bar U_i^{(n-1)}(s) 
\quad 
(i=1,2,3), 
\qquad 
U_4^{(n)}(s) \equiv U_4(s), 
\qquad 
U_\mu^{(0)}(s) \equiv U_\mu(s). 
\end{equation}

As an important feature, this smearing procedure keeps 
the gauge covariance of the smeared link-variable $U_\mu^{(n)}(s)$ properly. 
In fact, the gauge-transformation property of 
$U_\mu^{(n)}(s)$ is just the same as that of the original link-variable 
$U_\mu(s)$, and therefore 
the gauge invariance of $O(U_\mu^{(n)}(s))$ is ensured 
whenever $O(U_\mu (s))$ is gauge invariant. 

Since the smeared link-variable $U_\mu^{(n)}(s)$ includes a spatial 
extension, the ``line'' expressed with $U_\mu^{(n)}(s)$ physically 
corresponds to a ``flux tube'' with a spatial extension.
Therefore, if a suitable smearing is done, the ``line'' of the 
smeared link-variable is expected to be close to the ground-state flux tube. 
Here, the overlap between the ground-state operator and 
the 3Q-state operator at $t=0, T$ in the 3Q Wilson loop $W_{\rm 3Q}$   
is estimated by   
\begin{equation}
C_0 \equiv \langle W_{\rm 3Q} (T) \rangle ^{T+1} / 
\langle W_{\rm 3Q} (T+1) \rangle ^T \quad \in [0,1]. 
\end{equation}

To get the ground-state-dominant 3Q system, 
we investigate the ground-state overlap $C_0$ for  
the 3Q Wilson loop composed of the smeared link-variable $U_\mu^{(n)}(s)$ 
in the lattice QCD, and we adopt the smearing 
with $\alpha=2.3$ and the iteration number $n=12$, 
which largely enhances $C_0$ for all of the 3Q configurations 
in consideration as shown in Fig.2. 

%
%

\begin{figure}[htb]
\begin{minipage}[t]{60mm}
\epsfig{figure=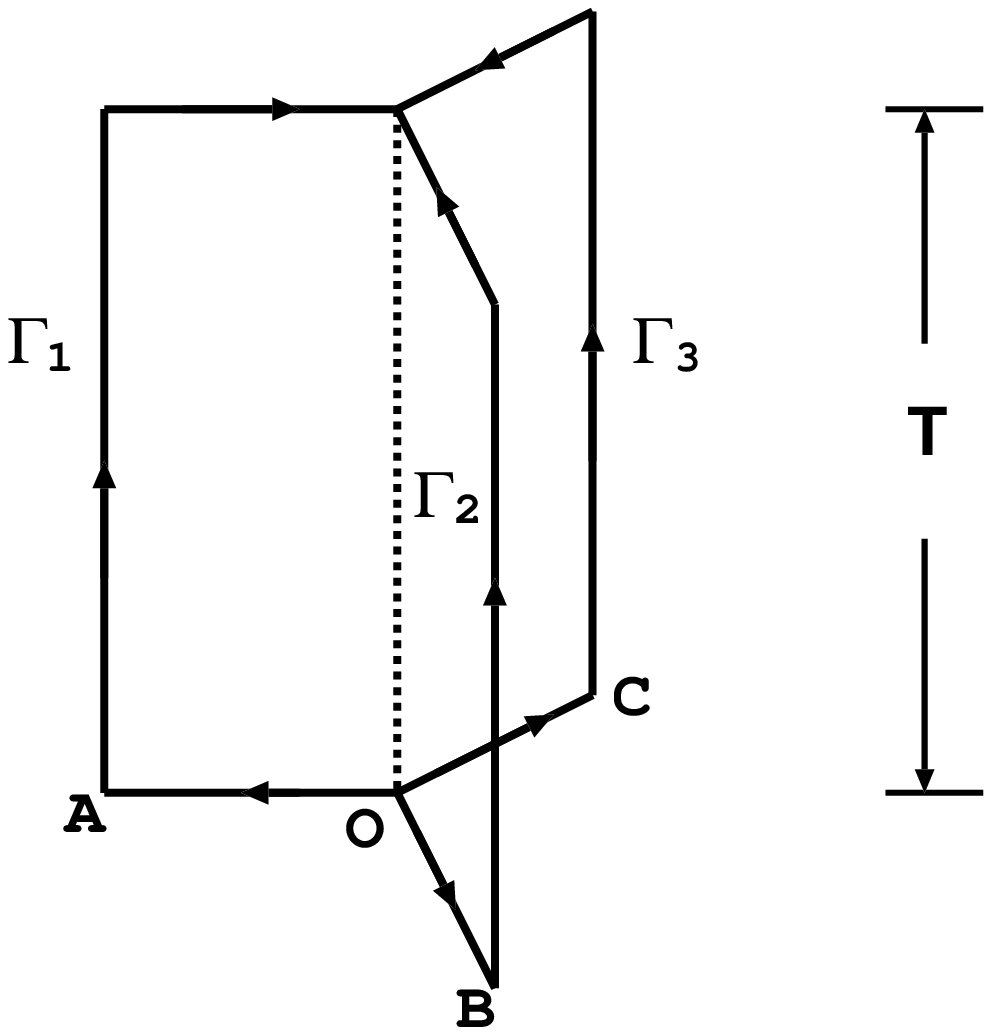, height=5.2cm}
\vspace{-1cm}
\caption{The 3Q Wilson loop $W_{\rm 3Q}$. 
The 3Q state is generated at $t=0$ and is annihilated at $t=T$.  
The three quarks are spatially fixed in ${\bf R}^3$ for $0 < t < T$.}
\end{minipage}
\hspace{\fill}
\begin{minipage}[t]{97mm}
\epsfig{figure=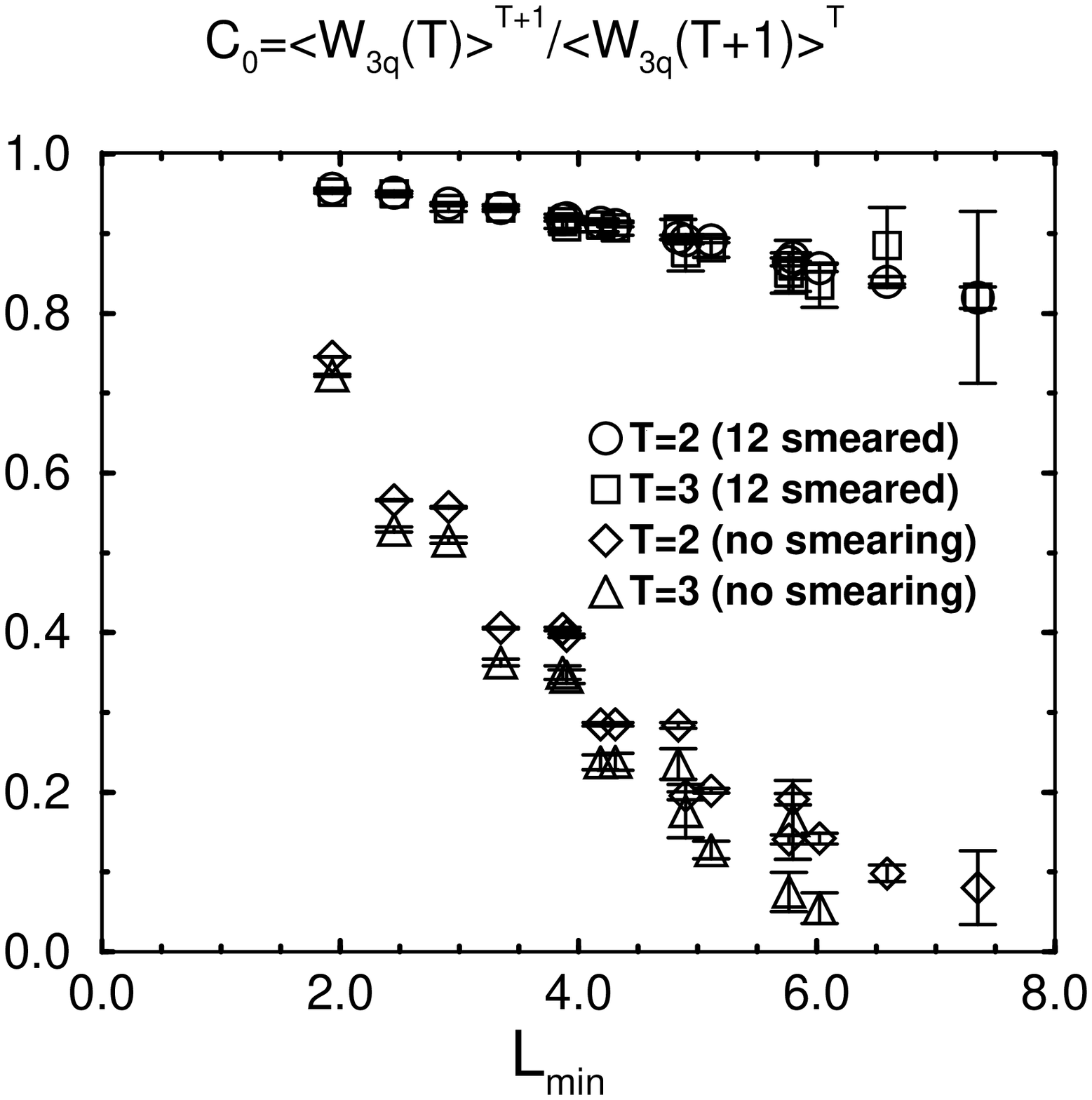, height=6cm}
\vspace{-1cm}
\caption{The ground-state overlap $C_0$ for the 3Q system 
with smeared link-variables (upper data) and with unsmeared 
link-variables (lower data). The horizontal axis denotes $L_{\rm min}$. 
For each 3Q configuration, 
$C_0$ is largely enhanced as $0.8 < C_0 < 1$ by the smearing.}
\end{minipage}
\end{figure}

\section{Theoretical Consideration for the 3Q Ground-State Potential}

In this section, we consider the potential form of $V_{\rm 3Q}$ based on QCD.  
In the short-distance limit, the perturbative QCD is applicable and 
the Coulomb-type potential appears as the one-gluon-exchange (OGE) result. 
In the long-distance limit at the quenched level, 
the flux-tube picture would be applicable from the argument 
of the strong-coupling QCD \cite{CI86,BPV95,KS75}, and hence 
a linear-type confinement potential is expected to appear.
%
Actually, the Q-$\bar{\rm Q}$ potential $V_{\rm Q \bar{Q}}$ 
is well reproduced by the Coulomb-plus-linear potential 
in the lattice QCD. 
Then, we conjecture that 
the 3Q ground-state potential $V_{\rm 3Q}$ is also expressed by 
a sum of the short-distance OGE result 
and the long-distance flux-tube result as 
\begin{equation}
V_{\rm 3Q}=-A_{\rm 3Q}\sum_{i<j}\frac1{|{\bf r}_i-{\bf r}_j|}
+\sigma_{\rm 3Q} L_{\rm min}+C_{\rm 3Q}, 
\quad 
V_{\rm Q \bar{Q}}(r)=-\frac{A_{\rm Q \bar{Q}}}{r}
+\sigma_{\rm Q \bar{Q}} r+C_{\rm Q \bar{Q}}, 
\end{equation}
where $L_{\rm min}$ denotes the minimal length of the total flux tubes  
linking the three quarks. 
%

Next, we consider the diquark limit, where 
the 3Q system becomes equivalent to a Q-$\bar{\rm Q}$ system, 
which leads to a physical requirement on the relation between 
$V_{\rm 3Q}$ and $V_{\rm Q \bar{Q}}$. 
Here, the constant term is to be considered carefully, 
because there appears a divergence from the Coulomb term 
in $V_{\rm 3Q}$ as 
$\frac{-A_{\rm 3Q}}{|{\bf r}_i-{\bf r}_j|} \rightarrow -\infty$  
in the continuum diquark limit. 
In the lattice regularization, this ultraviolet divergence 
is regularized to be a finite constant with the lattice spacing $a$ as 
$\frac{-A_{\rm 3Q}}{|{\bf r}_i-{\bf r}_j|} \rightarrow 
\frac{-A_{\rm 3Q}}{\omega a}$, where $\omega$ is a dimensionless constant 
satisfying $0 < \omega <1$ and $\omega \sim 1$. 
Thus, we get the diquark-limit requirement on the lattice, 
\vspace{-0.2cm}
\begin{equation}
\sigma_{\rm 3Q}=\sigma_{\rm Q\bar{Q}}, 
\qquad
A_{\rm 3Q}=\frac12 A_{\rm Q\bar{Q}}, 
\qquad
C_{\rm 3Q}-C_{\rm Q\bar{Q}}=\frac{A_{\rm 3Q}}{\omega a} \quad ( > 0). 
\end{equation}

\vspace{-0.5cm}
\section{Lattice QCD Results for the 3Q Ground-State Potential}

We measure the 3Q ground-state potential $V_{\rm 3Q}$ 
from 210 gauge configurations 
in the SU(3)$_c$ lattice QCD using the smearing technique, 
and compare the lattice data with the theoretical form of Eq.(5). 
Table 1 shows the best fit coefficients 
in $V_{\rm 3Q}$ and $V_{\rm Q \bar{Q}}$ in Eq.(5). 
Table 2 shows the comparison between the lattice QCD data 
$V_{\rm 3Q}^{\rm latt}$ and the fitting function $V_{\rm 3Q}^{\rm fit}$ 
in Eq.(5) with the three coefficients listed in Table 1. 

As shown in Table 2, the three-quark ground-state potential $V_{\rm 3Q}$ is 
well described by Eq.(5) with accuracy better than a few \%. 
From Table 1, we find a universal feature of the string tension, 
$\sigma_{\rm 3Q} \simeq \sigma_{\rm Q\bar{Q}}$, 
as well as the OGE result for the Coulomb coefficient, 
$A_{\rm 3Q} \simeq \frac12 A_{\rm Q\bar{Q}}$. 
All the diquark-limit requirements in Eq.(6) 
are satisfied for $\omega \simeq 0.46$. 


Another fitting with the $\Delta$-type flux-tube ansatz 
\cite{SW8486,TES88} seems rather worse, because of unacceptably large 
$\chi^2/N_{\rm DS} \ge 10.9$. 
However, as an approximation, $V_{\rm 3Q}$ seems described by a simple 
sum of an effective two-body Q-Q potential with a reduced string tension as 
$\sigma_{\rm QQ} \simeq 0.53  \sigma$. 
This reduction factor can be naturally understood as a geometrical factor 
rather than the color factor, since 
the ratio between $L_{\rm min}$ and the perimeter length $L_P$ satisfies 
$\frac12 \le \frac{L_{\rm min}}{L_P} \le \frac1{\sqrt{3}}$, 
which leads to $L_{\rm min} \sigma =L_P \sigma_{\rm QQ}$ 
with $\sigma_{\rm QQ}=(0.5 \sim 0.58) \sigma$. 

\begin{table}[htb]
\caption{
The string tension, the Coulomb coefficient and the constant term  
for the 3Q ground-state potential $V_{\rm 3Q}$ 
and the Q-$\bar {\rm Q}$ potential $V_{\rm Q\bar{Q}}$ in the lattice unit 
with $a \simeq$ 0.19fm.
}
\newcommand{\m}{\hphantom{$-$}}
\newcommand{\cc}[1]{\multicolumn{1}{c}{#1}}
\renewcommand{\tabcolsep}{2pc} 
\renewcommand{\arraystretch}{1.2} 
\begin{tabular}{@{}llll} \hline
                   & \cc{$\sigma$} & \cc{$A$}      & \cc{$C$}     \\ \hline
${\rm 3Q}$         & $0.1528(27)$  & $0.1316(62)$  & $0.9140(201)$  \\ 
${\rm Q\bar{Q}}$   & $0.1629(47)$  & $0.2793(116)$ & $0.6293(161)$ \\ 
\hline
\end{tabular}\\[2pt]
\end{table}

\vspace{-1.5cm}
\begin{table}[htb]
\caption{
The lattice QCD result for the 3Q ground-state potential  
$V_{\rm 3Q}^{\rm latt}$ and the fitting function 
$V_{\rm 3Q}^{\rm fit}$ in Eq.(5) 
for the 3Q system where the three quarks are put on $(i,0,0)$, 
$(0,j,0)$ and $(0,0,k)$ in ${\bf R}^3$ in the lattice unit. 
The error of $V_{\rm 3Q}^{\rm latt}$ denotes the statistical error 
estimated with the jackknife method. 
(The systematic error is not included.)  
}
\vspace{0.1cm}
\newcommand{\m}{\hphantom{$-$}}
\newcommand{\cc}[1]{\multicolumn{1}{c}{#1}}
\renewcommand{\tabcolsep}{2pc} 
\begin{tabular}{@{}llll}
\hline
$(i, j, k)$ & \cc{$V_{\rm 3Q}^{\rm latt}$} 
            & \cc{$V_{\rm 3Q}^{\rm fit}$} 
            & \cc{$V_{\rm 3Q}^{\rm latt}-V_{\rm 3Q}^{\rm fit}$} \\ 
\hline 
$(0, 1, 1)$ &  0.8459(36) &  0.8529 &   \m0.0070  \\ 
$(0, 1, 2)$ &  1.0970(40) &  1.1023 &   \m0.0053  \\ 
$(0, 1, 3)$ &  1.2935(39) &  1.2926 &  $-$0.0009  \\ 
$(0, 2, 2)$ &  1.3164(40) &  1.3262 &   \m0.0098  \\ 
$(0, 2, 3)$ &  1.5032(58) &  1.5069 &   \m0.0037  \\ 
$(0, 3, 3)$ &  1.6741(40) &  1.6808 &   \m0.0067  \\ 
$(1, 1, 1)$ &  1.0231(38) &  1.0091 &  $-$0.0140  \\ 
$(1, 1, 2)$ &  1.2181(61) &  1.2145 &  $-$0.0036  \\ 
$(1, 1, 3)$ &  1.4154(49) &  1.3958 &  $-$0.0196  \\ 
$(1, 2, 2)$ &  1.3870(46) &  1.3887 &   \m0.0017  \\ 
$(1, 2, 3)$ &  1.5588(60) &  1.5580 &  $-$0.0008  \\ 
$(1, 3, 3)$ &  1.7141(43) &  1.7195 &   \m0.0054  \\ 
$(2, 2, 2)$ &  1.5216(33) &  1.5230 &   \m0.0014  \\ 
$(2, 2, 3)$ &  1.6745(11) &  1.6755 &   \m0.0010  \\ 
$(2, 3, 3)$ &  1.8242(54) &  1.8169 &  $-$0.0073  \\ 
$(3, 3, 3)$ &  1.9607(92) &  1.9438 &  $-$0.0169  \\ 
\hline
\end{tabular}\\[2pt]
\end{table}

\end{document}